\documentclass[11pt,a4paper]{scrartcl}

\usepackage{CLICdp}

\usepackage{CLICdp_definitions}

\newcommand{\tabt}[1]{\multicolumn{1}{c}{#1}}
\newcommand{\tabtt}[1]{\multicolumn{2}{c}{#1}}

\newcommand{\mH }{\ensuremath{m_{\PH}}\xspace}
\newcommand{\GH }{\ensuremath{\Gamma_{\PH}}\xspace}
\newcommand{\BR}{\ensuremath{BR}\xspace}
\newcommand{\gHZZ}{\ensuremath{g_{\PH\PZ\PZ}}\xspace}
\newcommand{\gHWW}{\ensuremath{g_{\PH\PW\PW}}\xspace}
\newcommand{\gHtt}{\ensuremath{g_{\PH\PQt\PQt}}\xspace}
\newcommand{\gHbb}{\ensuremath{g_{\PH\PQb\PQb}}\xspace}
\newcommand{\gHcc}{\ensuremath{g_{\PH\PQc\PQc}}\xspace}
\newcommand{\gHTauTau}{\ensuremath{g_{\PH\PGt\PGt}}\xspace}
\newcommand{\gHMuMu}{\ensuremath{g_{\PH\PGm\PGm}}\xspace}

\title{CLIC Detector and Physics Status}

\clicdpconf{2017}{006}  
\date{\today}

\addauthor{N.~van der Kolk}{\institute{1}}
\addinstitute{1}{Max-Planck-Institut f\:ur Physik, Munich, Germany}

\onbehalfof{the CLICdp collaboration}

\abstract{
This contribution to LCWS2016 presents recent developments within the CLICdp collaboration.
An updated scenario for the staged operation of CLIC has been published; the accelerator will operate at \SI{380}{\GeV}, \SIlist{1.5; 3}{\TeV}. 
The lowest energy stage is optimised for precision Higgs and top physics, while the higher energy stages offer extended Higgs and BSM physics sensitivity.
The detector models CLIC\_SiD and CLIC\_ILD have been replaced by a single optimised detector; CLICdet.
Performance studies and R\&D in technologies to meet the requirements for this detector design are ongoing.
}

\titlecomment{Talk presented at the International Workshop on Future Linear Colliders (LCWS2016), Morioka, Japan, 5--9 December 2016. C16-12-05.4.}

\graphicspath{ {./Figures/} }

\begin{document}

\titlepage

\section{Introduction}
This contribution to LCWS2016 presents recent developments within the CLIC Detector and Physics Study Collaboration (CLICdp).
The CLICdp collaboration currently consists of 28 institutes from 17 countries.
Its main activities are detector R\&D and optimisation studies for a detector at the Compact LInear Collider (CLIC), as well as the study of physics prospects at CLIC using full detector simulations including pile-up from beam-induced backgrounds. 
Recently CLICdp has published two important milestone papers; an updated baseline scenario for the staged operation of CLIC~\cite{2016_Staging} and a Higgs physics overview paper~\cite{2016_Higgs}. Additionally a new detector concept has been defined~\cite{2017_CLICdet}.

\section{Updated CLIC staging baseline scenario}
The CLIC Conceptual Design Report (CDR)~\cite{2012_CDR} was published in 2012, before the discovery of the Higgs boson and as such the Higgs mass was not fully taken into account in the optimisation of the different energy stages.
Additionally, in the CDR the CLIC accelerator complex had been optimised for \SI{3} {TeV}, while the two lower energy stages at \SI{500} {GeV} and \SI{1.4/1.5} {TeV} were at that time not fully optimised.
Hence, after comprehensive studies of the CLIC performance, of cost and power optimisation and of further Higgs and top quark physics, an updated baseline staging scenario has been defined in which also the lower energy stages are optimised.

\begin{figure}[h]	
	\begin{center}		
		\includegraphics[width=0.7\textwidth]{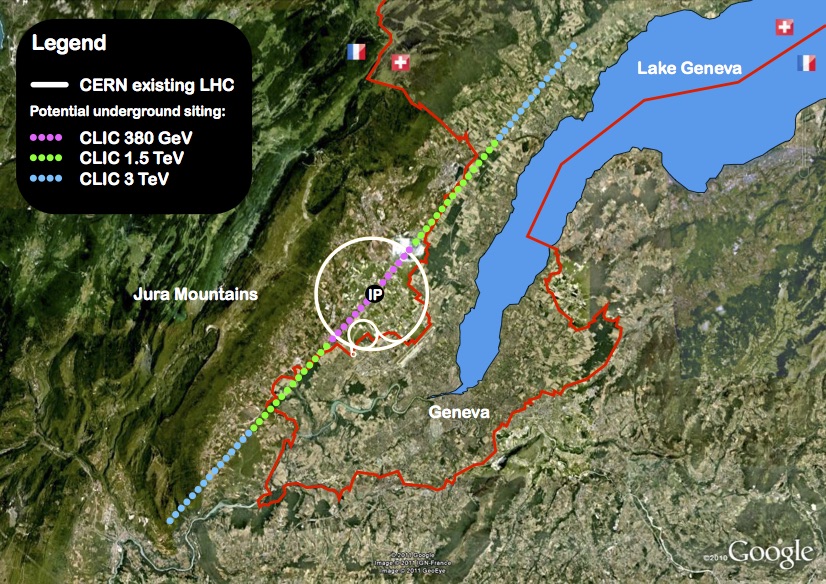}
		\caption{CLIC footprint in the Geneva area~\cite{2016_Staging}.}
		\label{fig:footprint}
	\end{center}
\end{figure}

In the new staging scenario CLIC will run at the following energies; \SI{380} {GeV}, \SIlist{1.5;3} {TeV}, collecting integrated luminosities of \SIlist{500; 1500; 3000} {\per \fb}, respectively. 
Additionally,  a total of \SI{100}{\per  \fb} of data will be collected in a threshold scan of energies around the top pair production threshold near \SI{350}{GeV}.
\cref{fig:footprint} shows the footprint of each of the CLIC energy stages in the vicinity of CERN.
The CLIC accelerator for \SI{380} {GeV} will be \SI{11.4} {\km} long and it will make use of one drive beam complex. The accelerating gradient will be lower than for the higher energy stages, namely \SI{72} {MV \per m}.
The higher energy stages of CLIC are defined based on the maximum energy attainable with the CLIC drive beam complex; with one complex this is \SI{1.5} {TeV} and with two \SI{3} {TeV}.
The accelerator will be \SI{29} {km} long for \SI{1.5} {TeV} and \SI{50} {km} for \SI{3} {TeV}. 
For these higher energy stages modules with an accelerating gradient of \SI{100} {MV \per m} will be used combined with the lower gradient modules of the initial \SI{380} {GeV} energy stage.

\begin{figure}[h]	
	\begin{subfigure}[b] {0.48\textwidth}		
		\includegraphics[width=\textwidth]{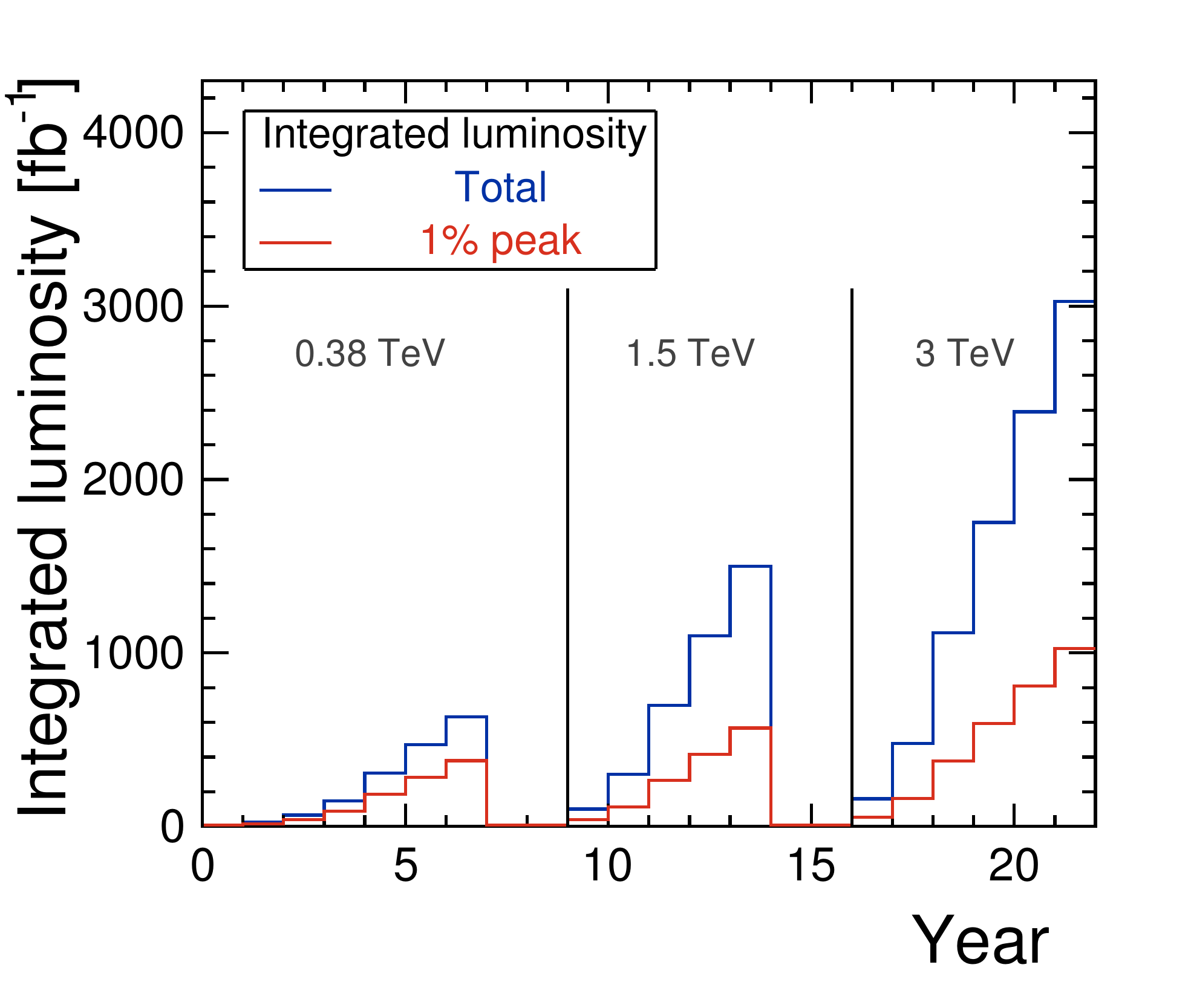}
		\caption{Integrated luminosity.}
		\label{fig:IntegratedLuminosity}
	\end{subfigure}
	\hfill
	\begin{subfigure}[b] {0.48\textwidth}	
		\includegraphics[width=\textwidth]{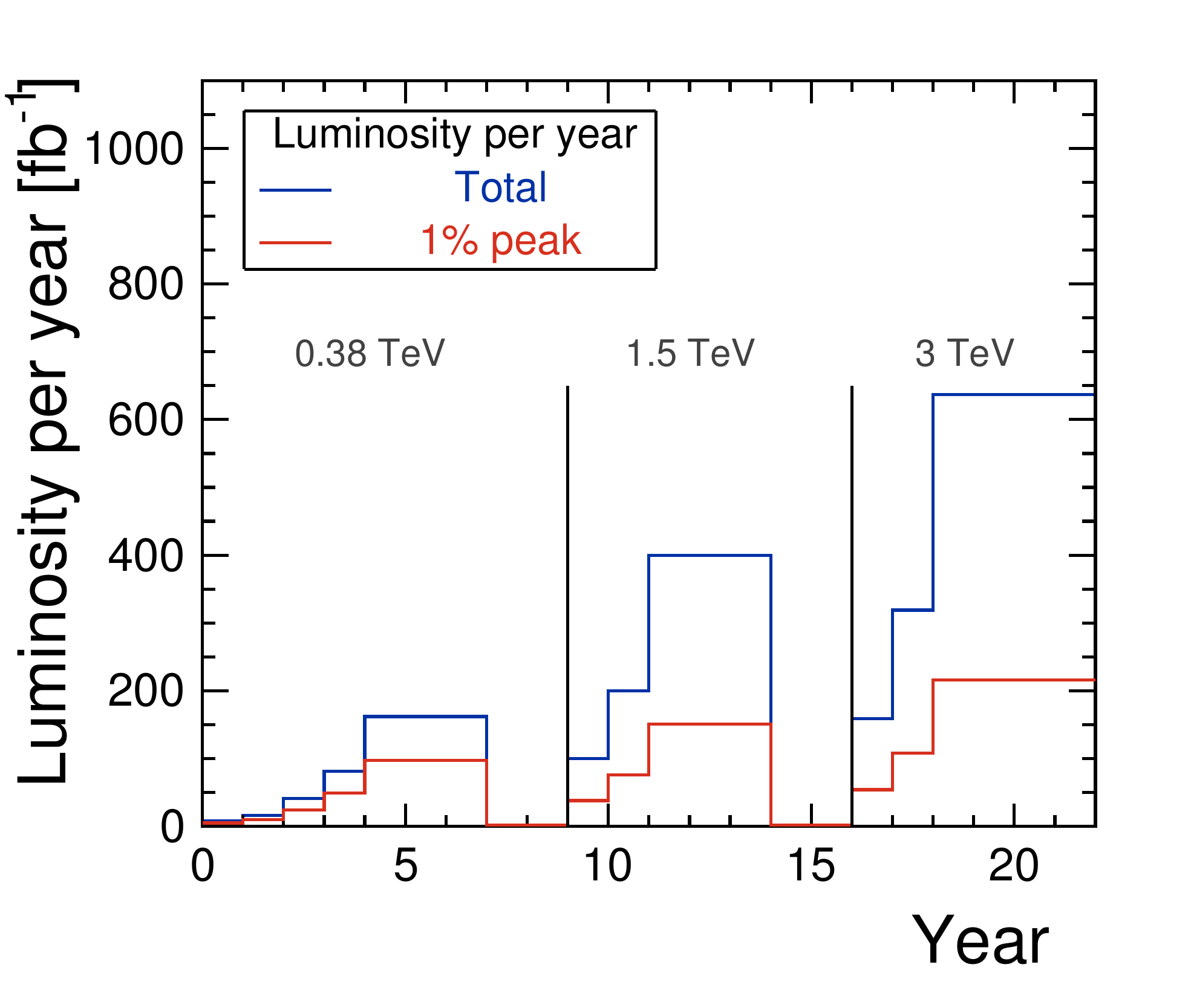}
		\caption{Luminosity per year.}
		\label{fig:LuminosityPerYear}
	\end{subfigure}
	\caption{Luminosity for the staged operation of CLIC, \subref{fig:IntegratedLuminosity} shows the integrated luminosity over the full CLIC running period and \subref{fig:LuminosityPerYear} shows the luminosity per year~\cite{2016_Staging}.}
\end{figure}

The full physics program will span 22 years, with 5 to 7 years of running at each stage. 
There will be upgrade periods of two years between the different energy stages. 
\cref{fig:IntegratedLuminosity} and \cref{fig:LuminosityPerYear} show the luminosity and the integrated luminosity for the full staging scenario.
For each stage the luminosity will be ramped up, as is illustrated in the figures.
The assumption is made that CLIC will operate for the equivalent of 125 full days per year, that is \num{1.08e7} seconds per year.
The motivations that led to the choice of the updated baseline staging scenario are described in detail in~\cite{2016_Staging}. 
A short summary is given below.

\subsection{Lowest energy stage at 380 GeV}
\begin{figure}[h]	
	\begin{minipage} [b] {0.48\textwidth}		
		\vspace{1cm}
		\includegraphics[height=0.29\textheight]{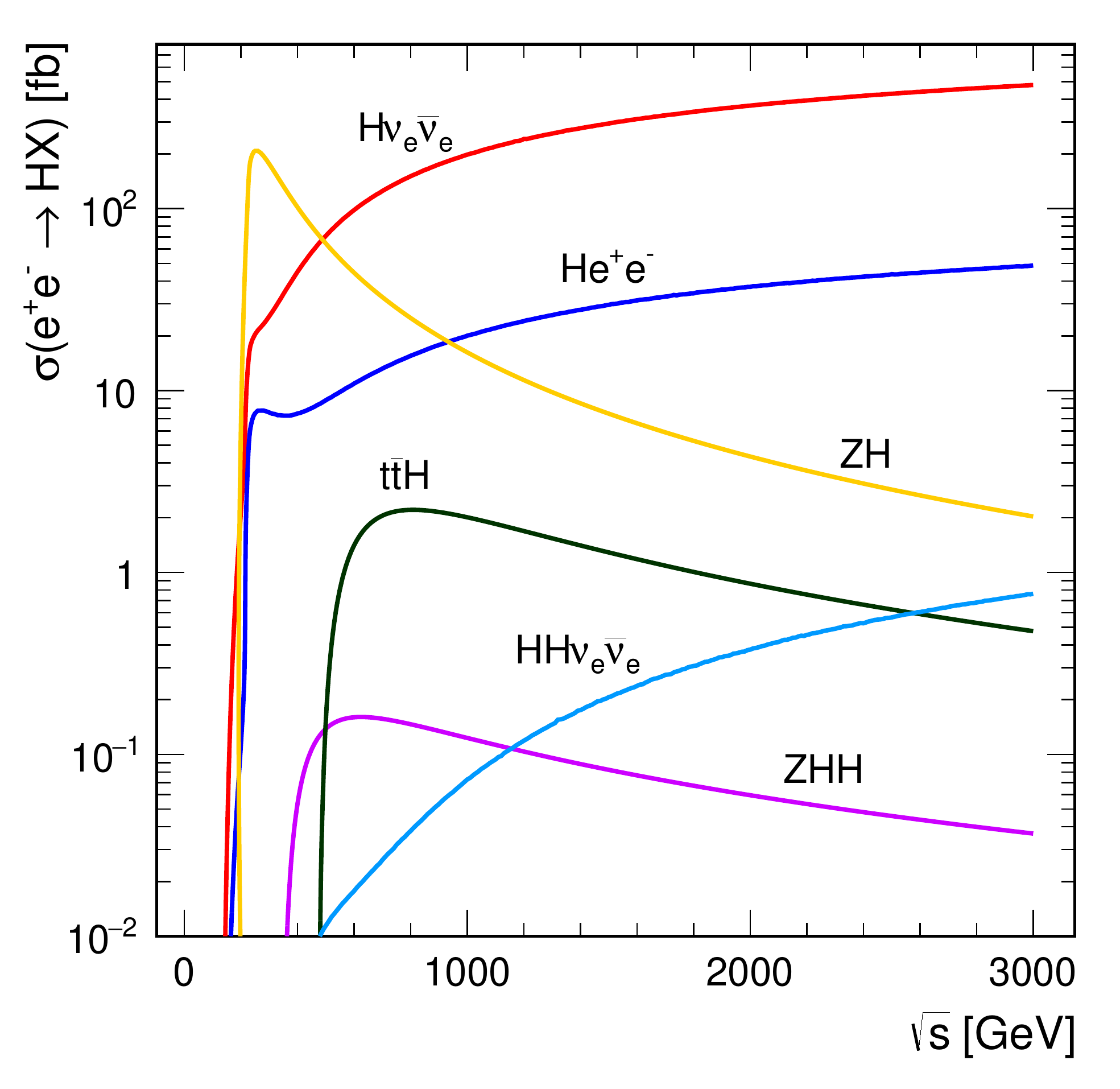}
		\vspace{2mm}
		\caption{Dominant Higgs production cross-sections at CLIC as a function of centre-of-mass collision energy~\cite{2016_Higgs}.}
		\label{fig:xsec}
	\end{minipage}
	\hfill
	\begin{minipage} [b] {0.48\textwidth}	
		\includegraphics[height=0.3\textheight]{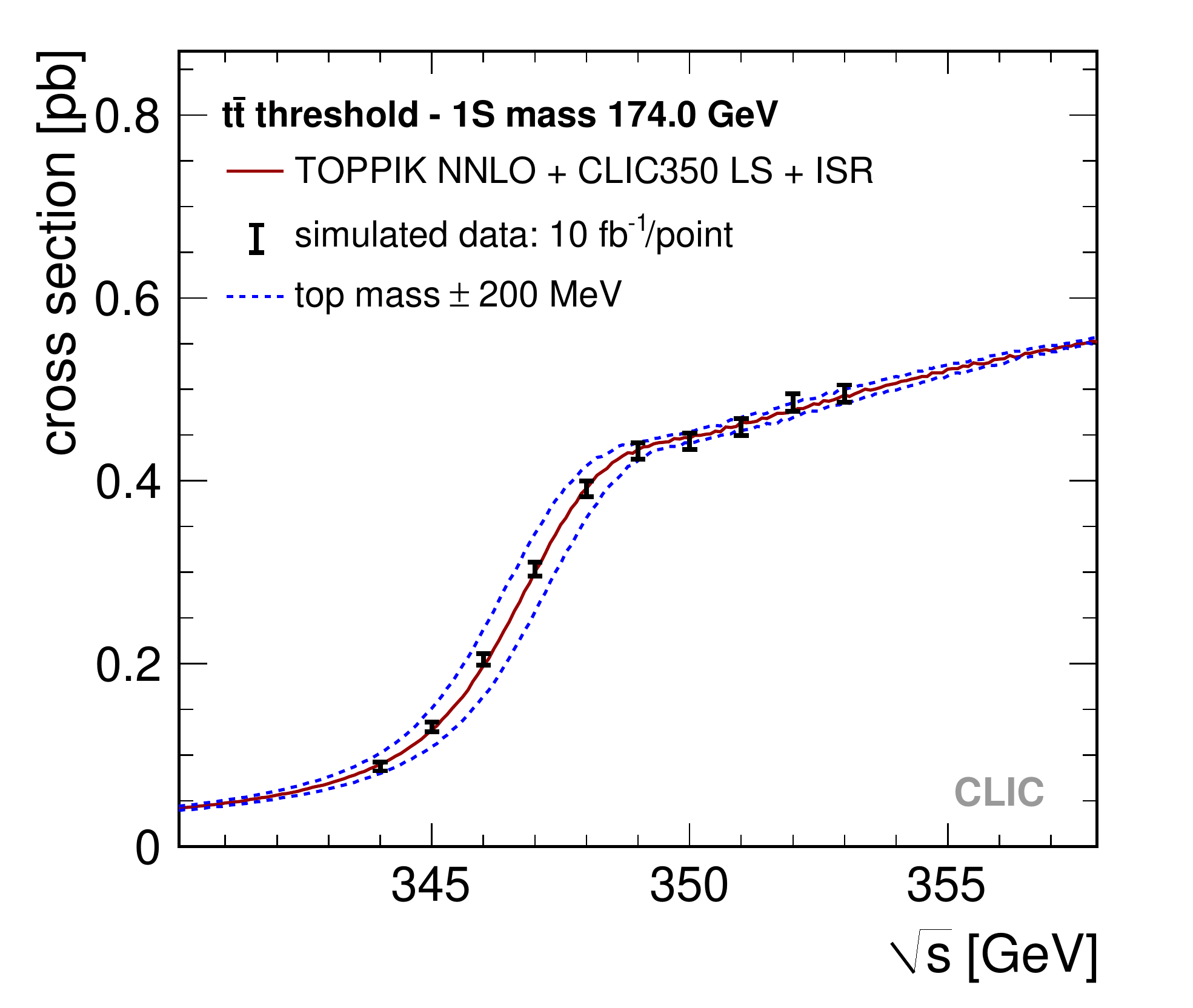}
		\caption{Top pair production cross section as a function of centre-of-mass energy~\cite{2013_Seidel}.\newline}
		\label{fig:TopThreshold}
	\end{minipage}
\end{figure}

The lowest energy stage of CLIC is aimed at precision Standard Model Higgs and top quark physics.
At this energy the Higgs mass can be determined with a statistical precision of about 100 MeV. 
The dominant Higgs production channel around this energy is Higgsstrahlung, $\Pep\Pem \rightarrow \PZ \PH$, as can be seen in \cref{fig:xsec}. 
This process can be used to measure the coupling of the Higgs to the \PZ, g$_{\PH\PZ\PZ}$, to a precision of 0.8\% at 350 GeV.
Combining this production channel with the WW-fusion channel, $\Pep\Pem \rightarrow \PH\PGne\PAGne$, gives access to the Higgs total decay width and to the Higgs coupling to the \PW, g$_{\PH\PW\PW}$, via the ratio of the cross-sections.
At lepton colliders a model independent determination of the Higgs couplings to fermions and bosons is possible and it depends on the absolute measurement of the Higgsstrahlung cross-section $\sigma(\PZ\PH)$. All other couplings are limited by the precision with which this coupling can be determined. 
Studies have shown that the best precision can be achieved around 350 GeV, despite the cross-section being higher around 250 GeV.
The total Higgs decay width can be found by combining the decay channels $\PH \rightarrow \PW\PW^{*}$ and $\PH \rightarrow \PZ\PZ^{*}$. 
Additionally, Higgs decays into invisible final states can be determined and used to constrain the invisible decay width of the Higgs.

\begin{figure}[h]	
	\begin{center}		
		\includegraphics[width=0.48\textwidth]{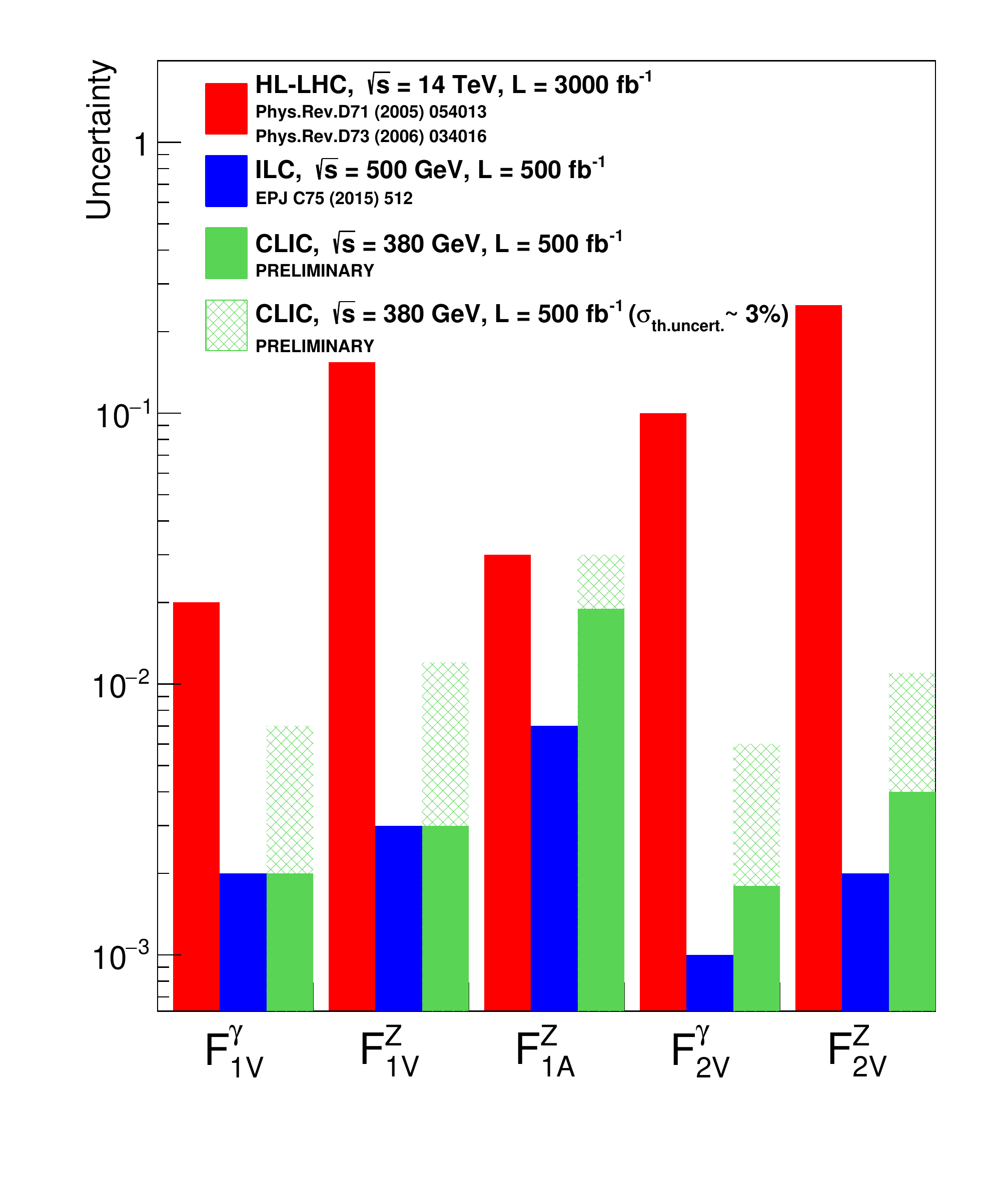}
		\caption{Uncertainties on the top quark form factors, comparing estimations from HL-LHC, ILC and CLIC~\cite{2016_Garcia}.}
		\label{fig:FormFactors}
	\end{center}
\end{figure}

The most precise measurement of the top quark mass can be obtained via a threshold scan around the top pair production threshold at around \SI{350} {GeV}, as illustrated in \cref{fig:TopThreshold}. 
A total uncertainty in the order of \SI{50} {\MeV} can be achieved~\cite{2016_Simon}. 
The top form factors can be accessed via the forward-backward asymmetry combined with cross-section measurements, and can be determined at the percent accuracy, as illustrated in \cref{fig:FormFactors}, where a comparison is made with HL-LHC and ILC.
The light green bars include an extra conservative theory uncertainty of \SI{3}{\percent}, as the exact calculations have not been done yet for \SI{380} {GeV}. 
Studies are being pursued to determine the form factors above the production threshold, where an increased boost leads to better separation between the decay products of the two top quarks.

The top quark is a promising candidate for the detection of physics processes that go beyond the Standard Model (BSM physics). 
BSM searches with a high statistical accuracy can be performed with top pair events recorded near the maximum of the top pair production cross-section at \SI{420} {GeV}. 
However, theory uncertainties are larger near the production threshold, and comparisons with theory would benefit from measurements at higher energies.

An energy of \SI{380} {\GeV} for the lowest energy CLIC stage follows from the physics described above; this energy is good for both Standard Model Higgs and top physics, as well as BSM physics.
At this stage \SI{15}{\percent} of the running time will be dedicated to a threshold scan of energies around the top pair production threshold, in order to determine the top mass with high precision.

\subsection{Higher energy stages}
\begin{figure}[h]		
	\begin{subfigure}[b] {0.48\textwidth}			
		\includegraphics[width=\textwidth]{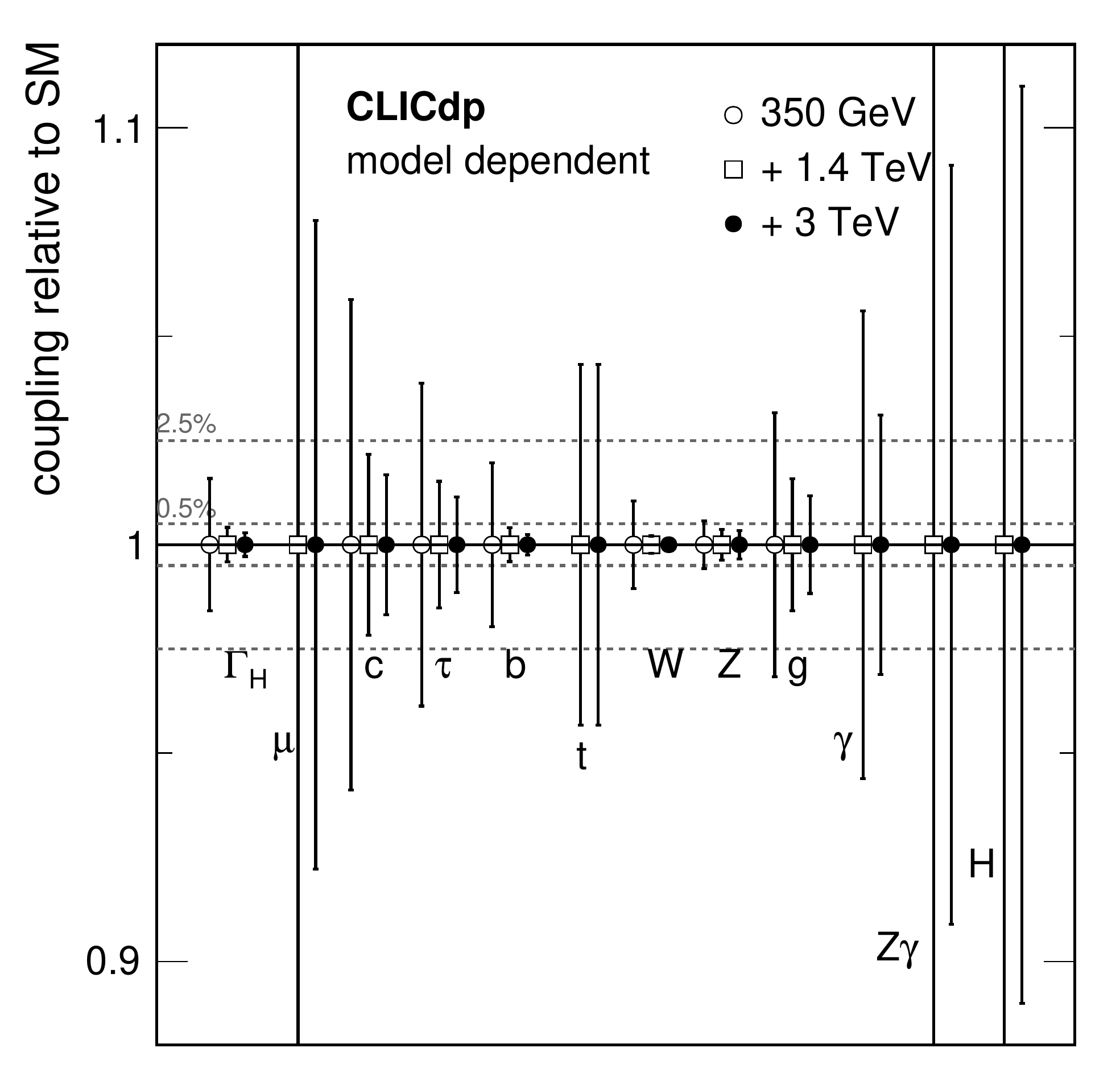}
		\caption{Model-dependent.}
		\label{fig:FitMD}
	\end{subfigure}
	\hfill
	\begin{subfigure}[b] {0.48\textwidth}	
		\includegraphics[width=\textwidth]{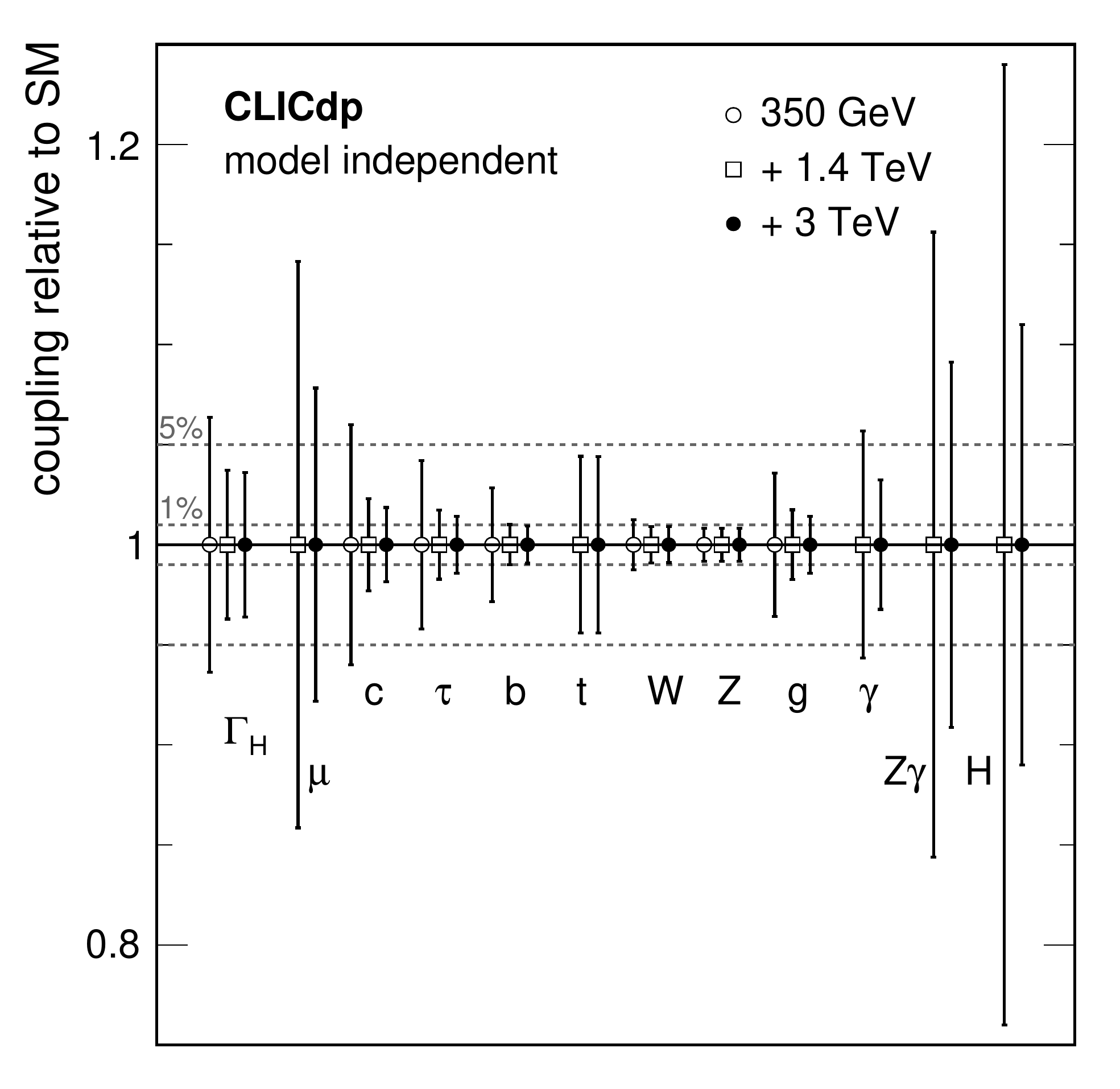}
		\caption{Model-independent.}
		\label{fig:FitMI}
	\end{subfigure}
	\caption{Precision of the Higgs couplings determined in \subref{fig:FitMD} a model-dependent fit and in \subref{fig:FitMI} a model-independent fit~\cite{2016_Staging}.}
\end{figure}

The higher energies offer extended potential in Higgs physics.
Above 1 TeV WW-fusion and ZZ-fusion are the dominant Higgs production channels. 
The high luminosities planned for these energies, combined with high cross-sections, enable the determination of the couplings of the Higgs to fermions and bosons at the \SI{1}{percent} level. 
However, this precision can only be reached with essential input from the \SI{380} {GeV} stage. 
The precision with which the couplings can be determined is illustrated in \cref{fig:FitMD} and \cref{fig:FitMI} for model-dependent and model-independent global fits to the data, respectively.
The Higgs mass can be determined to \SI{24} {MeV} through its decay $\PH \rightarrow \PQb\PAQb$ by combining data from \SIlist{1.5; 3} {TeV}. 
For this measurement polarisation of the electron beam helps to increase the accuracy.

Rare processes become available to study, e.g. the top Yukawa coupling can be determined from the process $\Pep\Pem \rightarrow \PQt \PAQt \PH$. 
Accuracies of \SIrange{4} {5} {\percent} can be achieved (depending on the electron beam polarisation).
Additionally, the Higgs tri-linear self-coupling $\lambda$ can be determined to within \SI{10}{\percent} at the HHH vertex via the rare process $\Pep\Pem \rightarrow \PH \PH \PGne \PAGne$, by combining the \SIlist{1.5; 3} {TeV} measurements.
The Higgs self-coupling gives direct access to the coupling parameter of the Higgs potential; it is therefore an essential measurement to establish experimentally the Standard Model Higgs mechanism.

\begin{figure}[h]	
	\begin{center}		
		\includegraphics[width=0.6\textwidth]{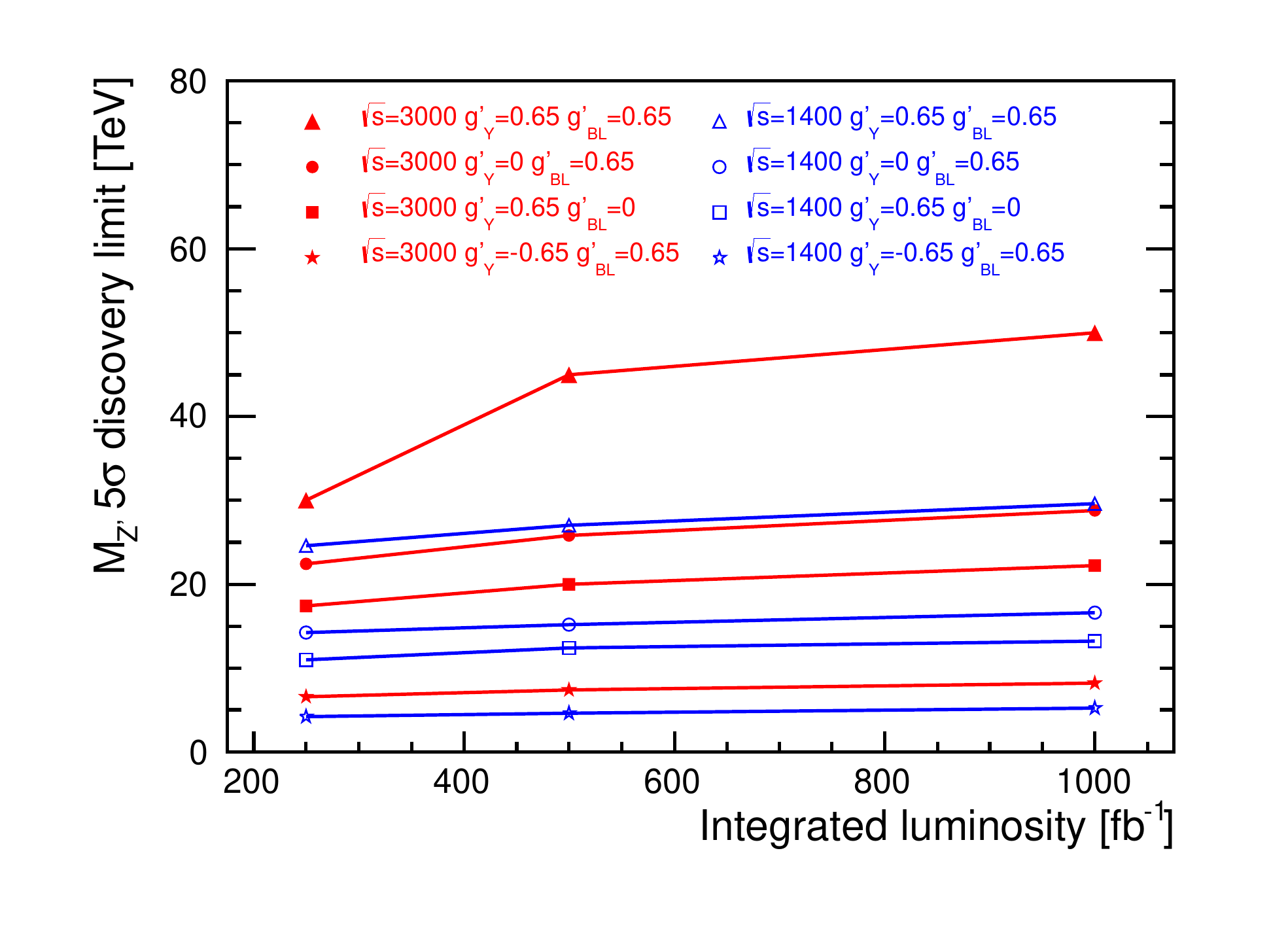}
		\caption{\PZpr mass discovery limit at 5$\sigma$ from the measurement of $\Pep\Pem \rightarrow \PGmp \PGmm$ for different model parameters as a function of the integrated luminosity~\cite{2012_Blaising, 2013_Abramowicz}.}
		\label{fig:ZpMass}
	\end{center}
\end{figure}

Higher energies additionally offer extended assess to BSM physics. 
Direct searches for new particles can be performed with a \SI{1}{\percent} accuracy on their mass measurement up to approximately half the centre-of-mass energy.
Additionally, for indirect searches of BSM physics CLIC reaches sensitivies beyond the centre-of-mass energy of the collider, through precision measurements of parameters and couplings in the Standard Model.
For example, the process $\Pep\Pem \rightarrow \PGmp \PGmm$ is sensitive to the presence of a high mass \PZpr boson. 
\cref{fig:ZpMass} illustrated the 5$\sigma$ discovery limit (in the minimal anomaly-free \PZpr model) up to tens of TeV for the \PZpr as a function of the integrated luminosity. 
In this study measurements of the cross-section, forward-backward asymmetry and the left-right asymmetry for opposite polarisations of the electron beam are used.
As already mentioned in the previous section, BSM physics in the top quark sector suffers from a lower cross section (reduced statistical accuracy) compared to measurements at \SI{380} {GeV}, but this is compensated by two effects: the top quark reconstruction efficiency is improved through the boost of the top quarks giving a better separation of the top quark decay products, and the relative BSM contribution in many BSM models is expected to increase with centre-of-mass energy.

\section{New CLIC detector model: CLICdet}
\begin{figure}[h]	
	\begin{minipage}[b]{0.48\textwidth}	
		\includegraphics[width=\textwidth]{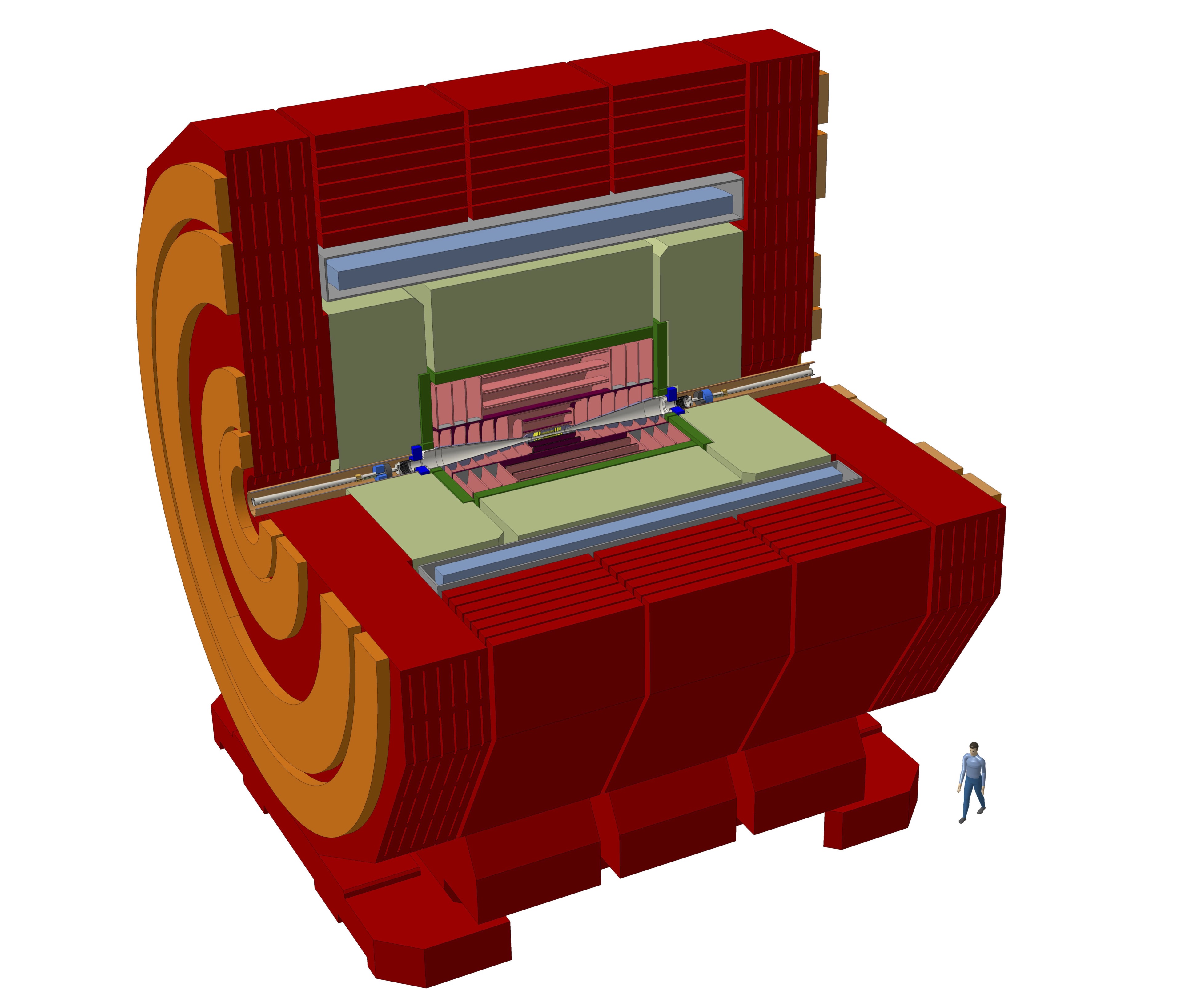}
		\caption{Illustration of the new single CLIC detector model CLICdet~\cite{2017_CLICdet}. Its overall height and length are \SIlist{12.9;11.4}{m}.}
		\label{fig:CLICdet}
	\end{minipage}
	\hfill
	\begin{minipage}[b]{0.48\textwidth}
		\includegraphics[width=\textwidth]{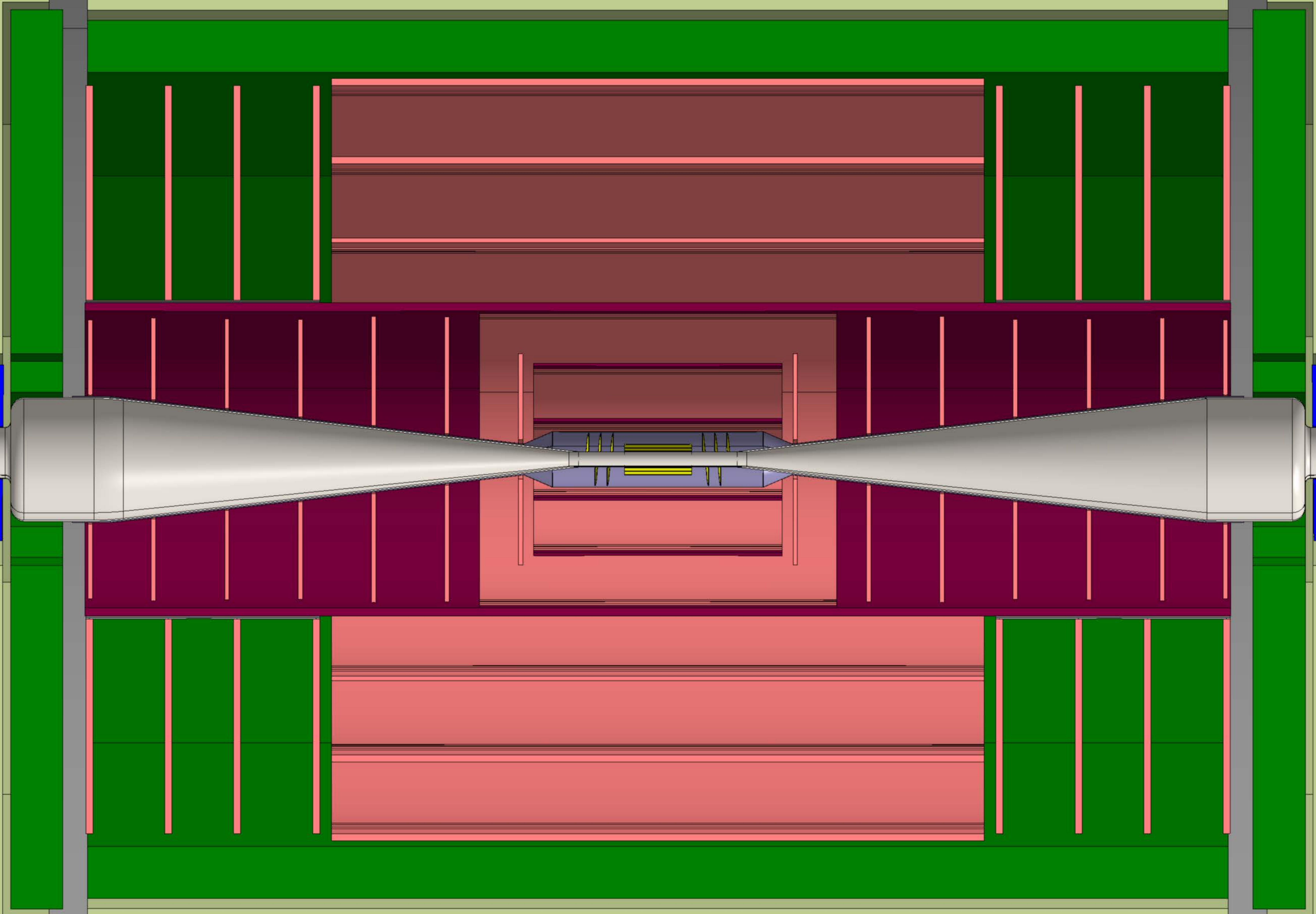}
		\vspace{5mm}
		\caption{Illustration of the tracker in CLICdet~\cite{2017_CLICdet}. Its overall height and length are \SIlist{3; 4.4} {m}, respectively.}
		\label{fig:tracker}
	\end{minipage}
\end{figure}

The CLIC CDR included two detectors concepts, each based on the detector models developed for the International Linear Collider, but adjusted to the CLIC environment: CLIC\_SiD and CLIC\_ILD. 
The beam-induced background levels at CLIC are particularly high in the multi-TeV region; there is a significant radiation of so-called beamstrahlung photons that leads to high rates of incoherent \Pep\Pem pairs and low p$_{\text{t}}$ $\PGg\PGg\rightarrow$hadrons events.
The energy lost through beamstrahlung generates a long lower energy tail in the luminosity spectrum.
The detector design is optimised to migitate the effects of the beam-induced backgrounds.
The two detector models are being used in the CLICdp physics studies.
Optimisation studies on these two models have led to a single new detector model for CLIC, called CLICdet, illustrated in \cref{fig:CLICdet}.
A detailed description of the new detector model is given in~\cite{2017_CLICdet}.
This model will be used in the next round of physics benchmark studies and has been fully implemented in the detector description toolkit DD4hep~\cite{dd4hep}.

The main differences of CLICdet with respect to the CDR detector design are the following;
there will be one single detector with an all silicon tracker and a magnetic field of \SI{4}{T},
the magnet return yoke has a smaller outer radius which is possible due to the less stringent requirements on the stray fields outside the magnet,
and the last quadrupole magnet, the QD0, will now be placed outside of the detector. 
This new position of the QD0, provides a much better forward coverage of the hadronic calorimeter. 
Forward coverage is very beneficial for reconstructing particles at low polar angles, a requirement in a number of important physics scenarios, especially at high energies. 
While in this configuration CLIC will operate with a larger L* of \SI{6} {m}, there is no significant loss in luminosity expected in the first energy stage, and for the higher energy stages a luminosity loss of only $\sim$\,\SI{10}{\percent} is estimated.


\subsection{Vertex and tracking detectors}
The CLICdet vertex detector consists of a cylindrical barrel closed off in the forward directions by disks. 
Unlike in the CDR, these disks are arranged to form a spiral geometry allowing efficient air-flow cooling. 
A spatial resolution of 3 $\mu$m, timing at the \SI{10} {ns} level, and a material budget of \SI{0.2}{\percent} of $X_{0}$ per layer is aimed for.
This requires novel sensor techniques which are being investigated in an extensive R\&D program.
These activities are closely linked to the R\&D for the silicon tracker.
For the vertex detector thin active edge sensors are being investigated, while for the tracker SOI chips are being developed in parallel to a CMOS test chip based on the one used for the ALICE experiment.
New larger chips are being designed, produced and tested and there have been systematic studies on capacitive coupling via finite element simulations, as well as tests of the alignment precision of glue assemblies.

\begin{figure}[h]	
	\begin{center}		
		\includegraphics[width=0.48\textwidth]{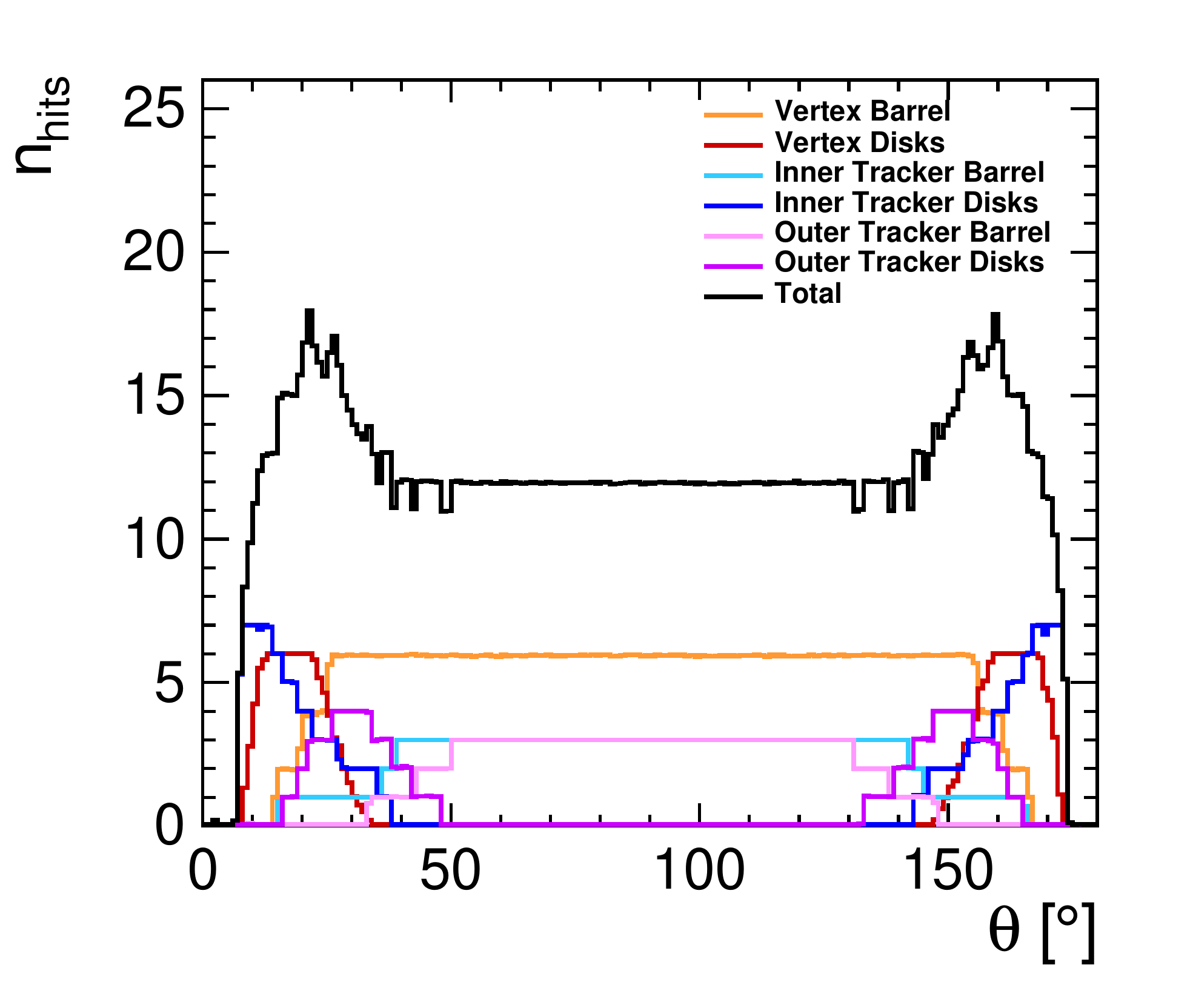}
		\caption{The coverage of the tracker system as a function of the polar angle. At least 8 hits are measured for muon tracks with a polar angle down to \SI{8}{\degree}~\cite{2017_CLICdet}.}
		\label{fig:Nhits}
	\end{center}
\end{figure}

The tracker is larger than CLIC\_SiD from the CDR. 
It offers 7 $\mu$m spatial resolution and \SI{10} {ns} timing information.
The new tracker layout is illustrated in \cref{fig:tracker}.
A larger tracker radius helps to improve the track momentum resolution, angular track resolution and jet energy resolution. 
Especially at the high energy stages an excellent tracker coverage also in the forward region is essential.
The position of the support tube, that divides the tracker into an inner and an outer part, is placed at a larger radius, enabling tracking very close to the beam pipe, as is demonstrated in \cref{fig:Nhits}, which shows the number of hits that are measured per muon track as a function of its polar angle. 
The inner part consists of 3 barrel layers and 7 disks, while the outer layers also consist of 3 barrel layers and 4 disks.
The largest challenge in the tracker design is to accommodate the large tracker volume with a material budget of only a little more than \SI{1}{\percent} of a radiation length per active layer.
This is potentially realised in the thin sensors and hybrid systems under study, but also requires R\&D on the support structure in order to achieve this low material budget.

\subsection{Calorimeters} 
\begin{figure}[h]	
	\begin{minipage}[b]{0.48\textwidth}
		\includegraphics[height=0.28\textheight]{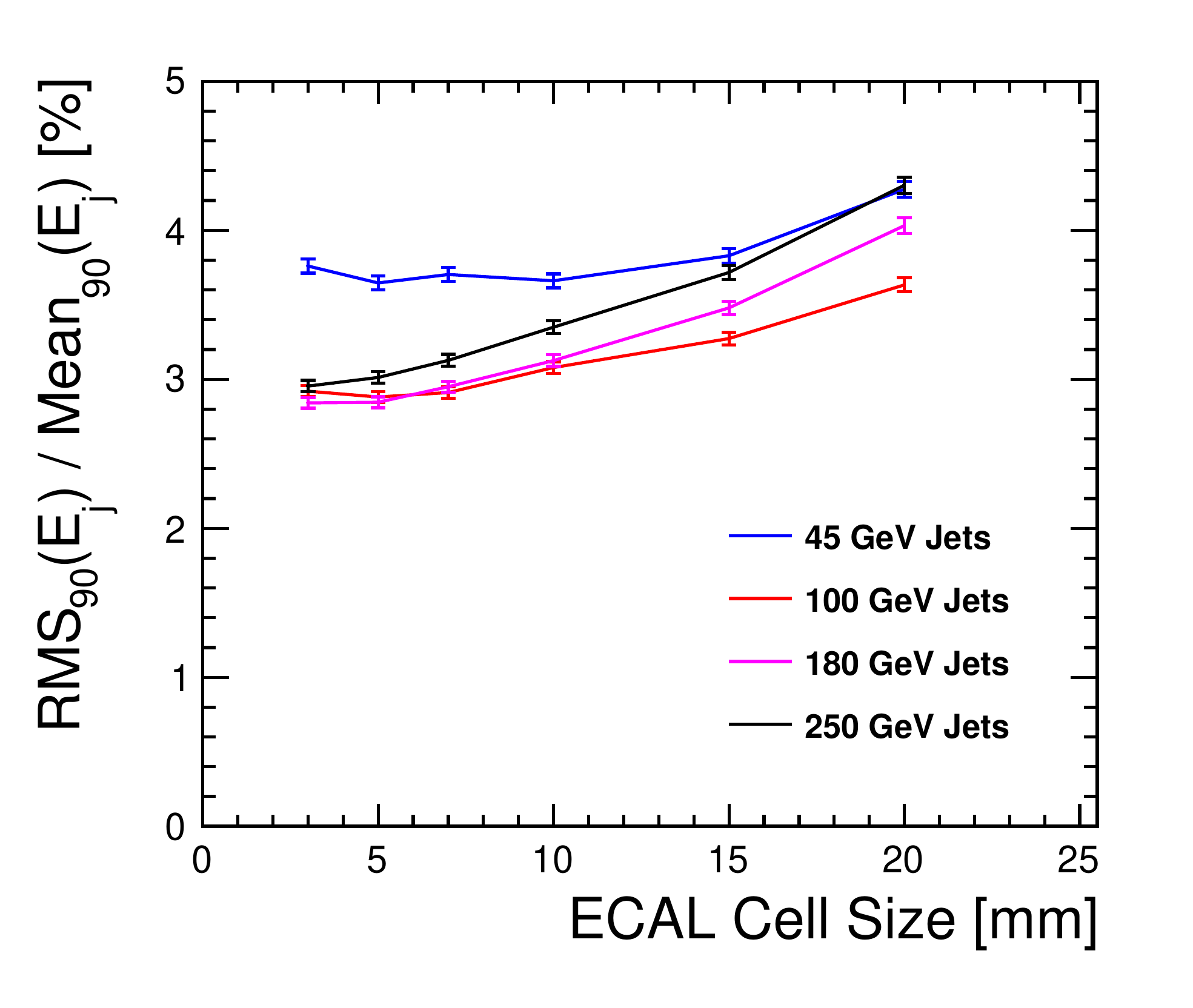}
		\caption{Energy resolution for jets of different energies, as a function of the SiW ECAL cell size~\cite{2017_CLICdet}.\newline}
		\label{fig:JERvsEcalCellSize}
	\end{minipage} %
	\hfill 
	\begin{minipage}[b]{0.48\textwidth}
		\includegraphics[height=0.25\textheight]{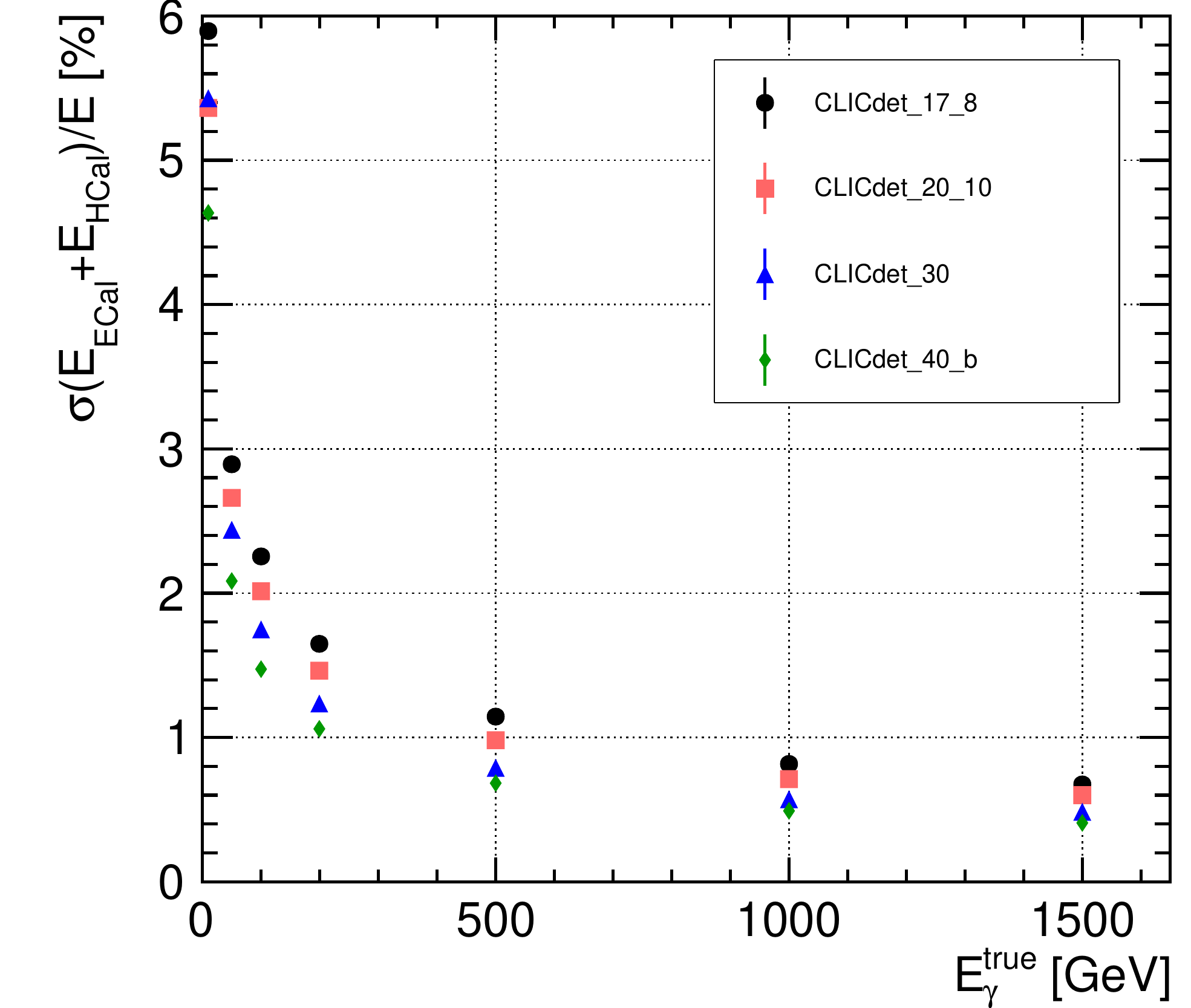}
		\vspace{3mm}
		\caption{Resolution for high energy photons of the total ECAL plus HCAL energy, for different ECAL models~\cite{2017_CLICdet}.\newline}
		\label{fig:EcalResolution}
	\end{minipage} 
\end{figure}

The electromagnetic calorimeter (ECAL) consists of 40 layers of tungsten absorbers interleaved with silicon sensors. 
It is based on a prototype from the CALICE Collaboration and has a high sensor granularity~\cite{2008_Repond}.
Its inner and outer radius are \SI{1.5}{m} and \SI{1.7} {m}, respectively.
The ECAL presented in the CDR was optimised considering the energy resolution of jets of particles. 
Since then the optimisation also takes into account the energy resolution of high energy photons, important in a large number of physics scenarios.
A cell size of \SI{5x5}{mm} is chosen as it is optimal for jet energy resolution, as can be seen in \cref{fig:JERvsEcalCellSize}.
Although the jet energy resolution does not depend strongly on the number of layers, the photon energy resolution does, and 40 layers of \SI{1.9} {mm} tungsten interleaved with silicon sensors are found to be optimal (see \cref{fig:EcalResolution}).
Note that the ECAL depth of 22\,$X_{0}$ is similar to the CDR design.

\begin{figure}[h]	
	\begin{center}		
		\includegraphics[width=0.48\textwidth]{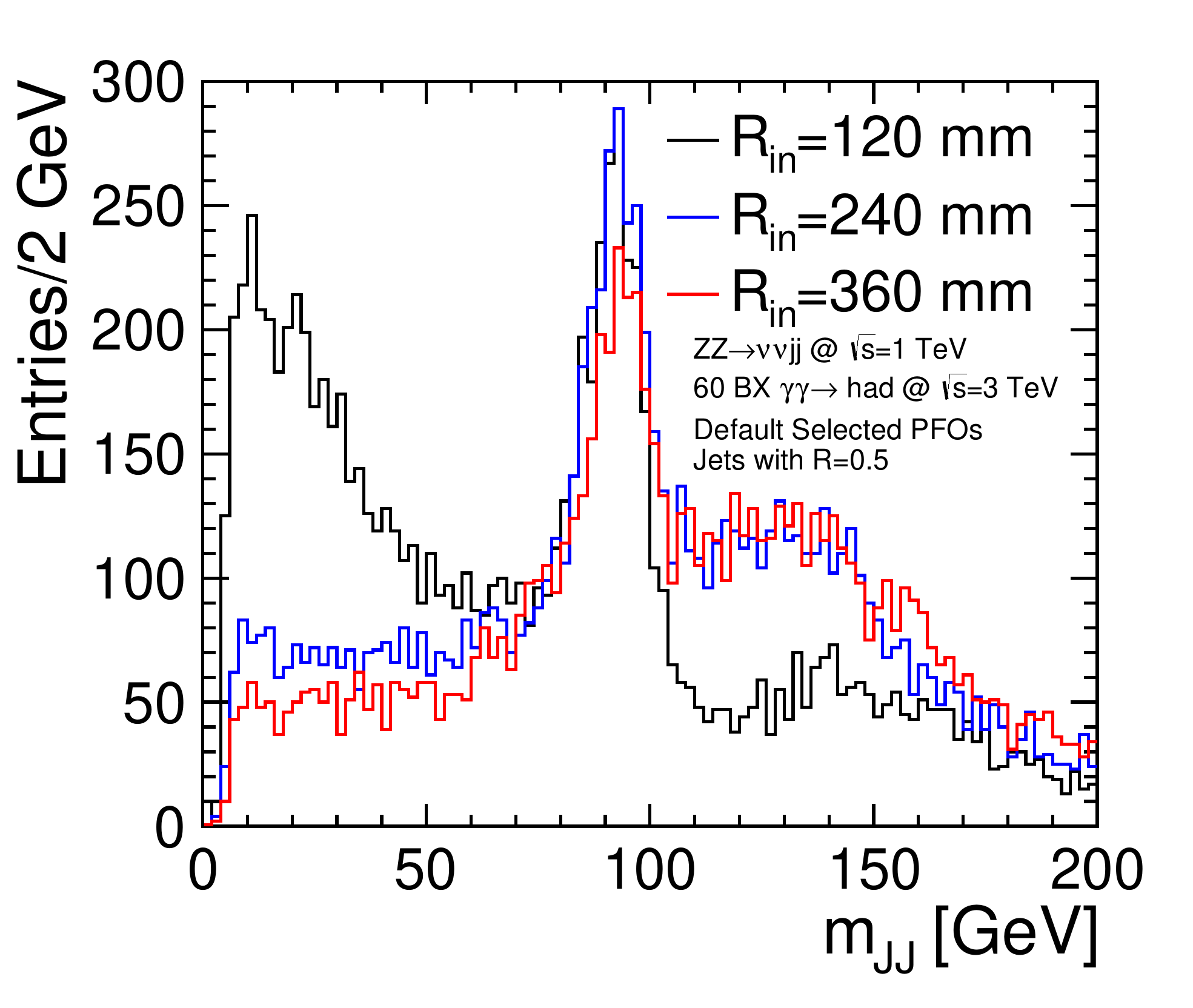}
		\caption{Di-jet invariant mass measurement for jets in the forward region in $\PZ\PZ \rightarrow \PGn \PGn \PQq \PQq$ events for different HCAL endcap radii~\cite{2017_CLICdet}.}
		\label{fig:Dijet}
	\end{center}
\end{figure}

\begin{figure}[h]	
	\begin{subfigure}[b] {0.48\textwidth}		
		\includegraphics[width=\textwidth]{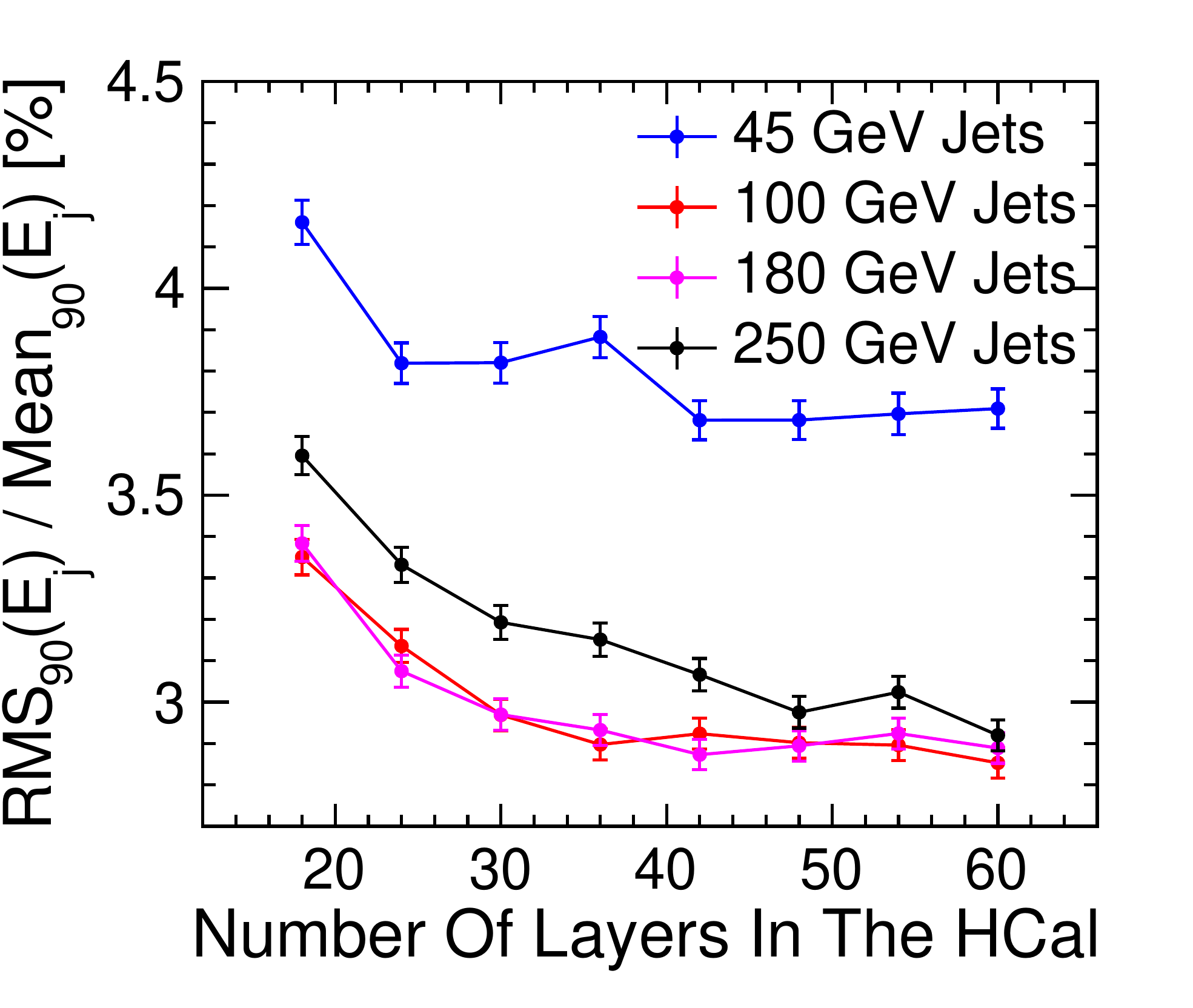}
		\caption{Number of HCAL layers.}
		\label{fig:JERvsLayers}
	\end{subfigure}
	\hfill
	\begin{subfigure}[b] {0.48\textwidth}	
		\includegraphics[width=\textwidth]{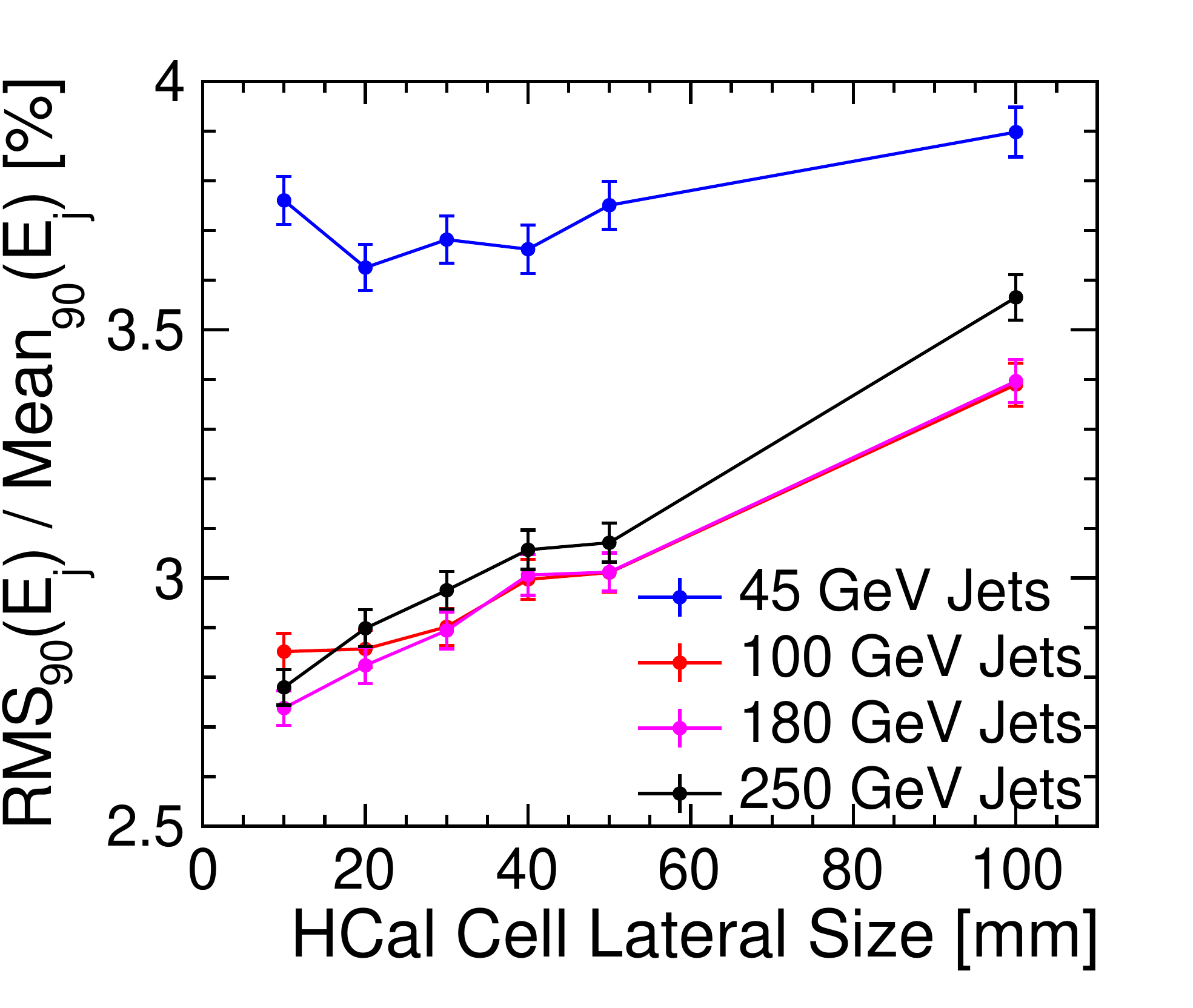}
		\caption{HCAL cell size. }
		\label{fig:JERvsHcalCellSize}
	\end{subfigure}
	\caption{Jet energy resolution for jets of different energies in the HCAL as a function of the number of HCAL layers in \subref{fig:JERvsLayers} and as a function of the HCAL cell size in \subref{fig:JERvsHcalCellSize}~\cite{2017_CLICdet}.}
\end{figure}

The hadronic calorimeter (HCAL) is designed to achieve a high jet energy resolution.
With respect to the CDR design the HCAL has a better forward coverage which improves the dijet invariant mass reconstruction as illustrated in \cref{fig:Dijet}.
The HCAL consists of 60 layers of \SI{19}{mm} thick steel absorbers interleaved with polystyrene scintillator tiles of \SI{30x30x3} {mm} individually read out by SiPMs, a design that also originates from the CALICE collaboration~\cite{2010_Adloff}.
The total HCAL depth is 7.5 interaction lengths.
Performance as a function of the number of layers and the cell size is shown in \cref{fig:JERvsLayers} and \cref{fig:JERvsHcalCellSize}, respectively.

The muon system integrated in the magnet return yoke is smaller compared to the CDR due to the thinner yoke in CLICdet, but it is still sufficient for muon identification.
There will be 6 sensitive layers, plus 1 additional layer in the barrel close to the magnet coil.
The current design consists of resistive plate chambers (RPC) with \SI{30x30} {mm} cell size, but alternatively scintillator strips are considered.

\begin{figure}[h]	
	\begin{center}		
		\includegraphics[width=0.6\textwidth]{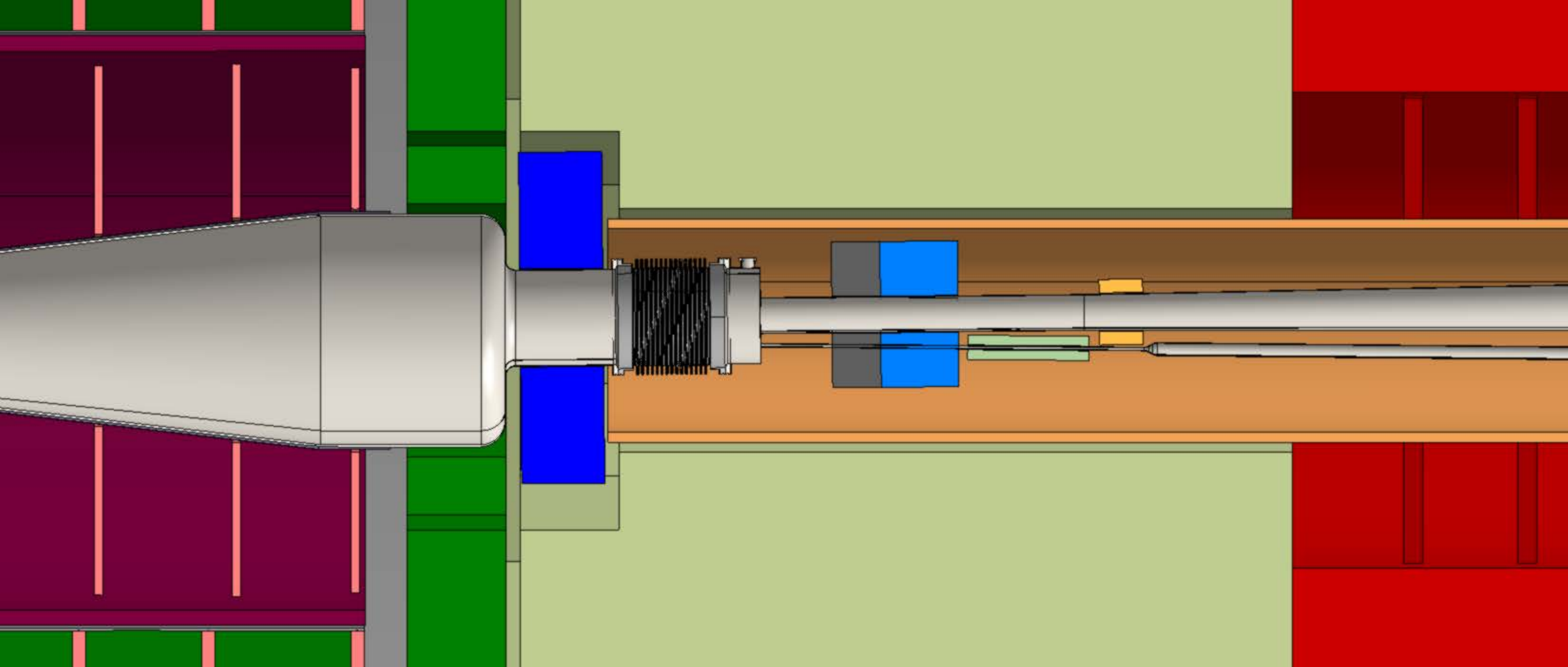}
		\caption{Layout of the forward region in CLICdet, seen from the top (LumiCal dark blue, BeamCal light blue)~\cite{2017_CLICdet}.}
		\label{fig:Forward}
	\end{center}
\end{figure}

The forward region of the detector houses additional calorimeters, LumiCal and BeamCal, originating from the FCAL collaboration~\cite{2010_Abramowicz}. 
Their position and radii, illustrated in \cref{fig:Forward}, have only slightly changed with respect to the CDR.
LumiCal (dark blue in the figure) will measure the luminosity via Bhabha scattering. 
It consists of 40 layers of tungsten and silicon sensors.
BeamCal (light blue in the figure) is installed for beam monitoring and also consists of 40 tungsten layers interleaved with a radiation hard sensor material (e.g. GaAs or diamond).
Additionally, these very forward calorimeters provide coverage for electrons and photons down to \SI{10}{mrad}.

\section{Further CLICdp activities}
Recently CLICdp has finalised a detailed Higgs physics overview paper~\cite{2016_Higgs}.
It demonstrates the CLIC Higgs physics reach at its 3 energy stages.
This paper combines more than 25 independent analyses in a collaboration wide effort. 
\cref{tab:GlobalFit:Input350} shows the summary of the precisions obtainable in the Higgs observables for the first stage of CLIC and \cref{tab:GlobalFit:Input143} shows the precision for the higher energy stages, where rare processes are also accessible.
The symbol '$-$' indicates that a measurement is not possible or relevant at this centre-of-mass energy, while the numbers marked with a '$^*$' were extrapolated from \SI{1.4}{TeV} to \SI{3}{TeV}.
The collaboration is now preparing similarly detailed papers outlining the CLIC potential for top quark physics, and for physics beyond the standard model, that will be finalised in 2017/2018.
      

\begin{table*}[htp]\centering
  \begin{tabular}{lllc}\toprule
                        &                                                           &                              & \tabt{Statistical precision}                        \\\cmidrule(l){4-4}
        \tabt{Channel}  & \tabt{Measurement}                                        & \tabt{Observable}            & \SI{350}{GeV}       \\ 
                        &                                                           &                              & $500\,\fbinv$        \\ \midrule
    $\PZ\PH$            & Recoil mass distribution                                  & $\mH$                        & \SI{110}{MeV}  \\
    $\PZ\PH$            & $\sigma(\PZ\PH)\times \BR(\PH\to\text{invisible})$         & $\Gamma_\text{inv}$          & $0.6\,\%$  \\ \midrule
    $\PZ\PH$            & $\sigma(\PZ\PH)\times \BR(\PZ\to\Plp\Plm)$             & $\gHZZ^{2}$                  & $3.8\,\%$  \\
    $\PZ\PH$            & $\sigma(\PZ\PH)\times \BR(\PZ\to\PQq\PAQq)$                  & $\gHZZ^{2}$                  & $1.8\,\%$  \\
    $\PZ\PH$            & $\sigma(\PZ\PH)\times \BR(\PH\to\PQb\PAQb)$                & $\gHZZ^{2}\gHbb^{2}/\GH$     & $0.84\,\%$ \\
    $\PZ\PH$            & $\sigma(\PZ\PH)\times \BR(\PH\to\PQc\PAQc)$                & $\gHZZ^{2}\gHcc^2/\GH$       & $10.3\,\%$ \\
    $\PZ\PH$            & $\sigma(\PZ\PH)\times \BR(\PH\to\Pg\Pg)$                   &                              & $4.5\,\%$ \\
    $\PZ\PH$            & $\sigma(\PZ\PH)\times \BR(\PH\to\tptm)$               & $\gHZZ^{2}\gHTauTau^{2}/\GH$ & $6.2\,\%$ \\
    $\PZ\PH$            & $\sigma(\PZ\PH)\times \BR(\PH\to\PW\PW^*)$                 & $\gHZZ^{2}\gHWW^{2}/\GH$     & $5.1\,\%$ \\
    $\PH\PGne\PAGne$    & $\sigma(\PH\PGne\PAGne)\times \BR(\PH\to\PQb\PAQb)$        & $\gHWW^{2}\gHbb^{2}/\GH$     & $1.9\,\%$ \\
    $\PH\PGne\PAGne$    & $\sigma(\PH\PGne\PAGne)\times \BR(\PH\to\PQc\PAQc)$        & $\gHWW^{2}\gHcc^{2}/\GH$     & $14.3\,\%$ \\
    $\PH\PGne\PAGne$    & $\sigma(\PH\PGne\PAGne)\times \BR(\PH\to\Pg\Pg)$        &     & $5.7\,\%$ \\    
    \bottomrule
  \end{tabular}
    \caption{Precision of the Higgs observables in the first stage of CLIC for an integrated luminosity of \SI{500}{\per \fb} at \SI{350}{GeV}, assuming
      unpolarised beams~\cite{2016_Higgs}.}      
      \label{tab:GlobalFit:Input350}
\end{table*}

\begin{table*}[htp]\centering
    \begin{tabular}{lllcc}\toprule
                        &                                                           &                              & \tabtt{Statistical precision}                        \\\cmidrule(l){4-5}
        \tabt{Channel}  & \tabt{Measurement}                                        & \tabt{Observable}  & \SI{1.4}{TeV}         & \SI{3}{TeV}           \\ 
                        &                                                           &                           & $1.5\,\abinv$      & $2.0\,\abinv$        \\ \midrule
   $\PH\PGne\PAGne$    & $\PH\to\PQb\PAQb$ mass distribution                       & $\mH$                & \SI{47}{MeV}     & \SI{44}{MeV}       \\ \midrule
   $\PH\PGne\PAGne$    & $\sigma(\PH\PGne\PAGne)\times \BR(\PH\to\PQb\PAQb)$        & $\gHWW^{2}\gHbb^{2}/\GH$   & $0.4\,\%$         & $0.3\,\%$           \\
   $\PH\PGne\PAGne$    & $\sigma(\PH\PGne\PAGne)\times \BR(\PH\to\PQc\PAQc)$        & $\gHWW^{2}\gHcc^{2}/\GH$  & $6.1\,\%$         & $6.9\,\%$           \\
   $\PH\PGne\PAGne$    & $\sigma(\PH\PGne\PAGne)\times \BR(\PH\to\Pg\Pg)$           &                     & $5.0\,\%$         & $4.3\,\%$           \\
   $\PH\PGne\PAGne$    & $\sigma(\PH\PGne\PAGne)\times \BR(\PH\to\tptm)$       & $\gHWW^{2}\gHTauTau^{2}/\GH$ & $4.2\,\%$         & $4.4\,\%$               \\
   $\PH\PGne\PAGne$    & $\sigma(\PH\PGne\PAGne)\times \BR(\PH\to\mpmm)$       & $\gHWW^{2}\gHMuMu^{2}/\GH$   & $38\,\%$        & $25\,\%$            \\
   $\PH\PGne\PAGne$    & $\sigma(\PH\PGne\PAGne)\times \BR(\PH\to\upgamma\upgamma)$ &                          & $15\,\%$          & $10\,\%^*$               \\
   $\PH\PGne\PAGne$    & $\sigma(\PH\PGne\PAGne)\times \BR(\PH\to\PZ\upgamma)$      &                             & $42\,\%$           & $30\,\%^*$               \\
   $\PH\PGne\PAGne$    & $\sigma(\PH\PGne\PAGne)\times \BR(\PH\to\PW\PW^*)$         & $\gHWW^{4}/\GH$            & $1.0\,\%$         & $0.7\,\%^*$         \\
   $\PH\PGne\PAGne$    & $\sigma(\PH\PGne\PAGne)\times \BR(\PH\to\PZ\PZ^*)$         & $\gHWW^{2}\gHZZ^{2}/\GH$  & $5.6\,\%$ & $3.9\,\%^*$   \\
   $\PH\epem$       & $\sigma(\PH\epem)\times \BR(\PH\to\PQb\PAQb)$           & $\gHZZ^{2}\gHbb^{2}/\GH$    & $1.8\,\%$ & $2.3\,\%^*$ \\ \midrule
   $\PQt\PAQt\PH$      & $\sigma(\PQt\PAQt\PH)\times \BR(\PH\to\PQb\PAQb)$          & $\gHtt^{2}\gHbb^{2}/\GH$  & $8.4\,\%$         & $-$             \\
   $\PH\PH\PGne\PAGne$ & $\sigma(\PH\PH\PGne\PAGne)$                               & $\lambda$                   & $32\,\%$          & $16\,\%$            \\
   $\PH\PH\PGne\PAGne$ & with $-80\,\%$ $\Pem$ polarisation                             & $\lambda$                  & $24\,\%$          & $12\,\%$            \\ \bottomrule
  \end{tabular}
  \caption{Precision of the Higgs observables in the higher-energy CLIC stages for integrated
    luminosities of $1.5\,\abinv$ at \SI{1.4}{TeV}, and $2.0\,\abinv$ at \SI{3}{TeV}. 
    In both cases unpolarised beams have been assumed~\cite{2016_Higgs}.}
    \label{tab:GlobalFit:Input143}
\end{table*}

In view of the next Update of the European Strategy for Particle Physics CLICdp is preparing a CLIC summary report based on the new staging scenario, the new detector model, CLIC technology R\&D, and the physics potential overview papers of Higgs, top and BSM physics.
This report aims to demonstrate that CLIC is an excellent option for a linear electron-positron collider and currently the only option for a multi-TeV lepton collider. 
If supported in the next European Strategy CLIC could be the next large collider facility at CERN.

\section{Conclusion}
After extensive studies of the physics potential, performance and cost, an updated, optimised staging scenario for CLIC has been defined.
At the lowest energy stage of \SI{380}{\GeV} CLIC offers a rich physics program of precision tests of Standard Model Higgs and top quark physics, including sensitivity to BSM physics.
The higher energy stages at \SIlist{1.5; 3} {TeV} offer extended Higgs and BSM physics sensitivity.
The potential for Higgs physics at CLIC has been extensively tested within the CLICdp collaboration and recently a paper summarising the full physics reach has been finalised. 
Similar extensive studies of top quark physics and BSM physics are ongoing.

A new optimised single detector model at CLIC has been developed following optimisation studies on the previous detector models.
A very active R\&D program for novel vertex and tracker technologies is being pursued within CLICdp.
Calorimeter R\&D is performed in close collaboration with the CALICE and FCAL collaborations.

The new detector model has been fully implemented in the simulation and reconstruction framework DD4hep and will be used for the next physics benchmark studies.
The full program of physics studies at CLIC will demonstrate that the staged operation of CLIC offers a physics programme that reaches far beyond the HL-LHC, making CLIC an excellent option for a next facility at CERN.

\printbibliography[title=References]

\end{document}